# The Green Premium Puzzle:
# Empirical Evidence from Climate-Friendly Food Products

Voraprapa Nakavachara[1]
Chanon Thongtai[1,+]
Thanarat Chalidabhongse[2]
Chanathip Pharino[3]

10 July 2025

Abstract:

This paper investigates whether climate-friendly food products command a price premium in consumer markets. Using product-level data from a supermarket in Sweden, we examine the relationship between front-of-package climate impact scores and retail prices, controlling for product size, nutritional content, and fixed effects. Contrary to the intuitive expectation of a positive green premium, we find no evidence that climate-friendly products are priced higher. In some product categories, products with better climate scores are in fact associated with lower prices, suggesting a negative premium—an outcome that gives rise to what we refer to as the *green premium puzzle*. We argue that market frictions—such as competing consumer priorities, psychological distance from climate issues, and skepticism toward environmental labeling—may suppress the price signals intended to reward sustainable consumption. These findings offer important insights for producers, retailers, and policymakers seeking to align climate goals with effective market incentives in the transition toward a more sustainable society.

[1] Faculty of Economics, Chulalongkorn University.
[2] Department of Computer Engineering, Faculty of Engineering, Chulalongkorn University.
[3] Department of Environmental Engineering, Faculty of Engineering, Chulalongkorn University.
[+] Corresponding Author's Email: Chanon.Th@chula.ac.th.
This Quick Win research project is supported by Ratchadapiseksompotch Fund Chulalongkorn University.

# 1. Introduction

Climate change is one of the most pressing challenges of our time, with post-industrial patterns of production and consumption driving widespread environmental degradation. Population growth and rising demand for food have placed increasing pressure on planetary resources, particularly in agriculture, leading to land degradation, water scarcity, pollution, and biodiversity loss (Islam & Zheng, 2024). In 2022 alone, 1.05 billion tons of food were wasted at the consumer level—across retail, food service, and households—contributing an estimated 8–10% of global greenhouse gas emissions (United Nations Environment Programme, 2024). These figures underscore the central role of food systems in shaping climate outcomes and the urgency of fostering more sustainable and responsible consumption practices.

Across the 27 EU member states, food systems contribute nearly 30% of total greenhouse gas emissions (Crippa et al., 2021), with meat and dairy products identified as among the most carbon-intensive food categories (Sinclair et al., 2025). At the same time, European consumers are among the most environmentally aware globally, and the region has taken a leadership role in advancing climate policy and sustainable consumption initiatives. Public concern over climate risks has helped shape regulatory frameworks that promote eco-labeling, sustainable procurement, and responsible market behavior (Schmidt, 2008).

In this context, Sweden, as an EU member, stands out for its strong institutional commitments and high levels of consumer engagement with sustainability. Sweden has committed to reducing greenhouse gas emissions by 40% by 2030 (Hof et al., 2016) and aims for 32% of gross final energy consumption from renewable sources (Jäger-Waldau et al., 2020). It was also one of the first countries to implement a carbon tax in 1991, with significant downstream effects on emissions and consumer behavior (Anderson, 2019). These national efforts reflect a broader commitment to transition toward climate-conscious economies and provide valuable contexts for exploring consumer responses to climate information.

In recent years, climate labeling schemes have emerged as a promising market-based strategy to encourage responsible consumption. By providing clear climate impact scores on food packaging, these labels aim to nudge consumer choices toward products with lower environmental footprints (Woerdman, 2004). Understanding the relationship between these climate scores and product pricing is critical: if low-emission products command a price premium, it may reflect strong demand for sustainability and reinforce incentives for innovation and eco-friendly production. If not—or if there is a negative premium—it may signal structural or perceptual barriers to responsible consumption, such as competing consumer priorities or distrust in environmental claims.

This paper uses the term *green premium* to refer specifically to price differences associated with food products that have lower life-cycle greenhouse gas emissions. While "green" can include a broad range of environmental characteristics, this study focuses on climate-related performance, as indicated by front-of-package climate scores.

This paper investigates whether climate-friendly food products command a price premium in consumer markets. Using real-world supermarket data from Sweden, we examine whether products with lower climate impact scores are priced higher or lower than their more carbon-intensive counterparts. To our knowledge, this is among the first studies to empirically assess the presence—or absence—of a climate-based green premium using observed market prices.

By analyzing consumer-facing climate information in Sweden, an environmentally advanced European country, we aim to contribute to the literature on responsible consumption and sustainability transitions. Our findings offer insights into how environmental information is reflected in real market prices—providing evidence on whether climate-friendly attributes are economically rewarded. This is valuable not only for academic understanding of sustainable consumption behavior but also for guiding businesses, policymakers, and sustainability advocates seeking to align market incentives with climate goals.

## 2. Literature Review and Theoretical Framework

The relationship between environmental product attributes and pricing is complex. Standard consumer theory suggests that products with superior attributes should command higher prices. In this case, better sustainability performances may be considered a superior or preferable attributes. Therfore, products with lower greenhouse gas emissions should command higher prices. This can result in a *green premium* that reflects consumers' willingness to pay for environmental benefits. These attributes are often communicated through tools like front-of-package climate scores.

However, real-world market behavior can deviate from this expectation. Various frictions—such as psychological distance from climate issues, competing consumer priorities, or skepticism toward environmental claims—may suppress demand for climate-friendly products or limit their ability to command a premium.

In essence, two opposing sets of market forces are at play: those that support the emergence of a green premium, and those that hinder or suppress it. This section reviews relevant theories and prior empirical studies to frame our investigation. Figure 1 presents the conceptual framework underlying our analysis.

**2.1 Market Forces Supporting a Green Premium**

2.1.1   Eco-Conscious Consumer preference

Environmental performance can serve as an added source of product value alongside traditional attributes such as quality and functionality (Getz & Page, 2015). Climate-conscious consumers often prefer products with higher environmental scores and are willing to pay more for them (Olson, 2013). Feucht and Zander (2018), using a mixed-methods approach—including choice experiments, surveys, and qualitative interviews—investigated carbon label effectiveness across six European countries (France, Germany, Italy, Norway, Spain, and the United Kingdom). Their findings show that carbon labels significantly increase purchase likelihood, with some consumers willing to pay a premium of up to 20% for labeled products.

2.1.2   Nudging via Climate Scores

Climate scores serve as a simplified decision-making tool (Thaler & Sunstein, 2021), enabling consumers to evaluate a product's environmental impact quickly without deep analysis. De-Loyde et al. (2022) suggest that pricing eco-labeled products higher than unlabeled ones can still be effective, as eco-labels strongly influence consumer choices—especially among those who can easily compare product attributes. Label design also plays a crucial role. Donato (2025) finds that consumers respond more favorably to eco-labels that are visually clean, conceptually straightforward, and use recognizable colors like green or blue. Feucht and Zander (2018) further recommend that effective carbon labels display a horizontal traffic-light color scale alongside absolute $CO_2$-equivalent values to improve consumer comprehension and trust.

### 2.1.3 Green Product Differentiation

Climate scores enable horizontal product differentiation, helping distinguish products along environmental dimensions even when they serve the same basic function (Jaiswal et al., 2021). Conrad (2005) argues that environmentally friendly attributes allow companies to occupy a distinct market position, often benefiting both green-oriented and conventional firms. Finisterra do Paço et al. (2009) find that green consumer segments differ based on both environmental and demographic factors. To effectively appeal to these segments, firms must tailor their positioning and communication strategies. Ignoring environmental preferences may result in reputational risks, particularly among sustainability-conscious consumers.

## 2.2 Market Forces Suppressing a Green Premium

### 2.2.1 Non-Locality and Psychological Distance

Environmental issues such as climate change often exhibit non-locality, meaning their causes and effects are dispersed across space and time. As a result, it is difficult for consumers to perceive who is directly affected by emissions from a particular product or activity (Yogeesh, 2024). This stands in contrast to more tangible and immediate issues such as animal welfare, where the consequences are visually evident and emotionally resonant (Watanabe, 2007). For example, Nakavachara et al. (2025) find a positive price premium for animal welfare-friendly products, suggesting that consumers are more responsive to visible and emotionally salient ethical attributes.

Because climate change effects are often distant and abstract, they may fail to trigger a comparable sense of urgency (Spence & Pidgeon, 2009). This psychological distance undermines the effectiveness of climate scores, which may feel less compelling compared to more personally relevant concerns (Lewandowsky et al., 2016). As a collective action problem rooted in structural injustice, climate change requires shared responsibility across consumers, firms, and institutions (Hormio, 2023).

### 2.2.2 Competing Priorities

The abstract nature of climate impacts also means that consumers may prioritize other product attributes over carbon footprint. For example, organic labels are often perceived as offering more concrete health benefits than climate labels and are therefore favored (Shaikh et al., 2024). Front-of-package labels frequently compete for attention—pesticide-free, nutrition score, bio-organic, animal-friendly, and fair trade certifications may all appear alongside climate scores (Cranfield & Magnusson, 2003). Consumers also weigh preferences such as taste: despite beef having approximately five times the carbon footprint of pork, many choose it based on flavor (Scholz et al., 2015). Rondoni and Grasso (2021) find that when carbon labels are displayed together with other certifications (e.g., organic or Fair Trade), consumers exhibit the lowest willingness to pay for carbon-related information.

### 2.2.3 Distrust and Confusion

Consumer perceptions of eco-labels are shaped by trust in the underlying information. While some consumers find climate labels helpful, others are skeptical, viewing them as marketing tools or potential greenwashing (Feucht & Zander, 2018). Gorton et al. (2021) show that institutional trust and third-party certification are essential in enhancing consumer confidence in eco-labels. However, the growing number of labels in the marketplace can lead to information overload and label fatigue. Moon et al. (2016) report that label similarity and proliferation can generate confusion, frustration, and negative emotions—reducing trust, satisfaction, and willingness to purchase. Miscommunication or lack of clarity around third-party verification can therefore undermine the credibility of climate labeling efforts.

**2.3 Supply-Side Innovation and the Porter Hypothesis**

While much of the green premium discussion focuses on consumer demand and market signals, supply-side factors may also explain why climate-friendly products do not always carry higher prices. According to the Porter Hypothesis (Porter & van der Linde, 1995), well-designed and well-enforced environmental regulations can stimulate innovation that improves firm productivity and competitiveness. These productivity gains can lead to increased market share or profitability—without necessarily requiring higher product prices.

Empirical studies also support this view. For example, Finger et al. (2019) found that precision agriculture technologies help reduce the use of inputs such as fertilizers and chemicals without affecting yields. This leads to lower costs for farmers while also reducing environmental impact. Rubashkina et al. (2015) find that environmental legislation is associated with a rise in patent applications for green technologies. These innovations not only promote emissions reductions but also generate spillover effects that improve efficiency in broader, non-environmental areas of production. From a firm's perspective, optimal pricing depends on a combination of chosen technology and cost structures (Mason, 2006). Efficiency gains from cleaner technologies can thus offset the costs of environmental compliance.

In this light, the absence of a price premium for low-emission products may reflect innovation-driven cost reductions rather than a lack of consumer willingness to pay. This supply-side perspective complements the demand-side frictions discussed earlier, offering a more holistic understanding of the green premium puzzle.

The theoretical frameworks discussed above reveal various market forces that can shape the relationship between climate scores and product prices. These competing forces can potentially result in positive premiums, negligible differences, or even negative premiums—depending on which forces are more dominant. While it may seem intuitive to expect a positive green premium, given the rising environmental awareness and sustainability goals among consumers and firms, the reality is more nuanced. As reviewed, both demand-side frictions and supply-side dynamics can suppress price premiums, even when environmental performance improves. Therefore, this paper relies on empirical analysis to help determine which set of forces—those supporting or those suppressing a green premium—ultimately dominates in real market settings.

## 3. Data and Methodology

**3.1 Dat**

We utilize data from a Swedish supermarket (as of March 2025).[1] We include all food categories for which sufficient information is available.[2] To ensure consistency in the measurement unit for product size, we retain only products with the "size" variable specified in grams. Sizes originally recorded in kilograms are converted to grams, while products measured by volume (e.g., liters) are excluded. Therefore, we drop the entire beverage category and also other items whose sizes are reported in volume rather than weight. The final sample consists of 5,458. Summary statistics for food products from the Swedish supermarket are presented in Table 1.[3,4]

---

[1] The data on the website are all in Swedish and we used google translate to obtain the English information.
[2] We had to dropped "Fish & Seafood", "Fruit & Vegetables", and "Vegetarian" categories due to insufficient information.

Table 1 shows that the logarithm of price ranges from 0.92 to 7.60, with a mean of 3.48. Product package sizes range from 0.22 grams to 4,500 grams, with an average of 357.93 grams. In terms of nutritional attributes, the average product contains 101.47 grams of carbohydrates, 39.80 grams of fat, 31.80 grams of protein, 4.57 grams of salt, and provides 911.20 kcal of energy.

The climate score is based on the product's carbon footprint, calculated as emission intensity—i.e., the amount of greenhouse gas emissions (converted to carbon dioxide equivalents) per kilogram of product. The supermarket sourced these data from the RISE Food Climate Database, which provides life cycle assessments (LCA) of Swedish food products. Products are then assigned a climate score based on their emission intensity as follows:

- Category 1 (Best): 0–0.5 kg $CO_2$eq/kg
- Category 2: >0.5–3 kg $CO_2$eq/kg
- Category 3: >3–10 kg $CO_2$eq/kg
- Category 4: >10–20 kg $CO_2$eq/kg
- Category 5 (Worst): >20 kg $CO_2$eq/kg

In this original scheme, a score of 1 indicates the lowest (i.e., most favorable) emission intensity, and 5 indicates the highest (least favorable).

---

[3] Products with logarithm of price less than zero were also dropped.
[4] Note that not all products have climate impact scores. Eventually, in the empirical analyses, products with no climate impact scores will be dropped in the regressions.

To enhance interpretability, we invert the climate score so that a value of 5 represents the best environmental performance (i.e., lowest emission intensity), and 1 represents the worst (i.e., highest emission intensity). This transformation aligns the scale with the intuitive notion that "higher is bettter" and facilitates a more straightforward interpretation within our econometric framework. After adjusting the scores, the average climate score is 3.89 and range from 1 to 5, with 5 being the most favorable.

Looking at product categories, the supermarket employs a three-level product category hierarchy. Table 2 presents the breakdown of Level 1 product categories—the broadest level. The supermarket categorizes food products into the following Level 1 categories: Bread & Bakery; Candy, Ice Cream & Snacks; Cheese; Dairy & Eggs; Delicacies; Freezer; Larder; Meat, Poultry & Charcuterie; Ready Meals & Snacks; and Spices & Seasonings. The full details of the respective category structures are provided in Appendix 1

**3.2 Methodology**

In this paper, we aim to investigate whether climate-friendly food products are associated with a positive green premium—or conversely, whether they are priced lower than their less sustainable counterparts. Our empirical analysis seeks to identify whether products with more favorable climate impact scores command higher prices or, instead, face a price penalty. The direction of the price effect may depend on a variety of factors, such as consumers' willingness to pay for sustainability, trust in climate labeling, or market frictions that prevent green preferences from being fully reflected in prices. Specifically, we estimate the following econometric model:

$$\ln(Price_i) = \beta_0 + \beta_1 Climate_i + \beta_2' \mathbf{X}_i + \theta_{country} + \gamma_{CategoryL3} + \varepsilon_i \qquad (1)$$

Here, the dependent variable is the natural logarithm of the price for product *i*. The key explanatory variable, *Climate*$_i$, captures the product's climate impact rating. The control vector $\mathbf{X_i}$ includes observable product characteristics such as package size and nutritional values (carbohydrates, fat, protein, salt, and energy content). To control for unobserved heterogeneity, we include fixed effects for the product's country of origin ($\theta_{country}$) and for detailed product categories at level 3 ($\gamma_{CategoryL3}$). We cluster standard errors at the Level 3 category level to correct for intra-cluster dependence in pricing outcomes.

The empirical analyses will be conducted using the Swedish supermarket data. We will begin by analyzing the full dataset, followed by separate analyses for products sourced from within the European market and those from outside the European market.[5] In addition, we will conduct separate analyses for each Level 1 product category to examine potential similarity and variation across product types.

As discussed earlier, the theoretical framework suggests that competing forces shape the presence and direction of the green (climate) premium. On one hand, certain forces are expected to drive a positive premium. These include eco-conscious consumer preferences, nudging effects through climate scores, and green product differentiation. On the other hand, several market frictions may suppress the premium or even lead to a negative one. These include non-locality and psychological distance, competing consumer priorities, and distrust or confusion around environmental claims.

---

[5] The term 'European market' refers to the geographic continent of Europe in this context.

Since the theoretical framework allows for opposing outcomes depending on which forces dominate, empirical analysis is necessary to uncover the actual pricing patterns. In this paper, we explore these dynamics using data from a Swedish supermarket—a case we believe to be indicative of broader trends in European markets, where consumers are relatively informed and engaged with climate issues. By focusing on a country where product-level sustainability information is both accessible and standardized, we aim to provide insights into the functioning of climate-conscious consumption in advanced economies. These findings may offer useful benchmarks and serve to inspire similar analyses elsewhere.

## 4. Results and Discussion

**4.1 Main Results**

Table 3 displays the main regression results for the Swedish dataset. Column (1) presents estimates using the full sample, while Columns (2) and (3) restrict the analysis to products manufactured within Europe and outside of Europe, respectively.

In Column (1), the coefficient on the *Climate* variable is statistically insignificant, suggesting no evidence of a green premium in the overall sample. This may imply that the positive and negative market forces influencing price offset each other. However, when the sample is limited to products produced within Europe (Column 2), the *Climate* variable becomes negative and statistically significant at the 10% level. A one-point improvement in the climate score is associated with an 8.68% reduction in price, indicating a negative premium for climate-friendly products. In contrast, for products manufactured outside of Europe (Column 3), the coefficient is again insignificant, implying no discernible climate-related pricing pattern in this subgroup.

Across most specifications, product size is positively associated with price—larger packages tend to cost more. In some models, higher fat or protein content is also linked to higher prices.

Tables 4 presents the regression results disaggregated by product category at Level 1. Most categories show statistically insignificant coefficients on the *Climate* variable, indicating no strong evidence of either a green premium or a price penalty. However, three categories—Freezer; Meat, Poultry & Charcuterie; and Ready Meals & Snacks—exhibit significant negative coefficients. In these cases, a one-point improvement in the climate score is associated with price decreases of 11.6%, 22.1%, and 5.3%, respectively, suggesting that climate-friendly products in these segments are priced lower.

**4.2 Robustness Tests**

As a robustness check, we conduct additional analyses using a different data source—product-level data from a Swiss supermarket (as of February 2025). It is important to note that this data source may follow slightly different methodologies. For instance, while the Swiss supermarket applies similar principles in assigning climate scores, the classification thresholds differ slightly. The climate score is assigned as follows:

- 1 Star (Worst): >10 kg $CO_2$eq/kg
- 2 Stars: 4.8–9.9 kg $CO_2$eq/kg
- 3 Stars: 2–4.7 kg $CO_2$eq/kg
- 4 Stars: 1–1.9 kg $CO_2$eq/kg
- 5 Stars (Best): 0–0.9 kg $CO_2$eq/kg

In addition, although the Swiss supermarket also uses a three-level product category hierarchy, its categorization methods differ from the Swedish dataset and cannot be fully harmonized. The purpose of including this dataset is to provide a robustness analysis that supports the generalizability of our findings. The summary statistics, Level 1 category breakdown, and full category details of the Swiss supermarket data are presented in Appendices 2, 3, and 4, respectively.

Table 5 displays the main regression results for the Swiss dataset. Column (1) presents estimates using the full sample, while Columns (2) and (3) restrict the analysis to products manufactured within Europe and outside of Europe, respectively. The results in Column (1) reveal a statistically significant negative relationship between climate score and price at the 1% level. Specifically, a one-point increase in the climate score leads to a 5.49% decrease in price. This negative premium remains robust in the within-Europe subsample (Column 2), with a similar magnitude: a 5.53% price reduction per one-point climate score improvement, also significant at the 1% level. However, in the outside-Europe subsample (Column 3), the climate score is no longer statistically significant.

Tables 6 presents the regression results disaggregated by product category at Level 1. Coefficients for the *Climate* variable are either insignificant or negative and statistically significant, again indicating no premium or a negative premium. Significant negative coefficients are found for the categories Bread, Pastries & Breakfast; Frozen Food; Meat & Fish; and Pasta, Condiments & Canned Food, with associated price reductions of 9.03%, 9.38%, 5.89%, and 6.70%, respectively, for each one-point increase in the climate score.

The results from the Swiss supermarket support our finding in the main analyses (Swedish data) that climate-friendly products do not command a positive price premium. In fact, both datasets consistently show evidence of a negative premium, suggesting that products with better climate scores tend to be priced lower rather than higher. This reinforces the robustness of our conclusion and highlights the green premium puzzle in climate-friendly food consumption.

**4.3 Discussion**

Reflecting on the results alongside the theoretical framework, which posits competing forces behind the green premium, we observe that the empirical findings lean toward one side. As outlined earlier, several mechanisms could support a positive premium for climate-friendly products—such as eco-conscious consumer preferences, the nudging effect of climate scores, and product differentiation based on environmental performance. At the same time, various market frictions—including non-locality and psychological distance, competing priorities, and skepticism or confusion regarding environmental claims—can suppress or even reverse the premium.

While the theoretical framework does not yield a definitive prediction, the empirical evidence from both the Swedish and Swiss supermarket datasets suggests that negative forces currently outweigh positive ones. In our Swedish analysis, a one-point improvement in climate score was associated with a statistically significant 8.68% price decrease, with particularly strong effects in specific product categories such as Freezer (–11.6%), Meat, Poultry & Charcuterie (–22.1%), and Ready Meals & Snacks (–5.30%). Similarly, the Swiss dataset revealed a 5.5% price reduction per one-point improvement, with sizable negative effects observed in categories such as Bread, Pastries & Breakfast (–9.03%), Frozen Food (–9.38%), Meat & Fish (–5.89%), and Pasta,

Condiments & Canned Food (–6.70%). Across both datasets, climate-friendly products were consistently associated with lower, not higher, prices.

These findings may appear counterintuitive, especially in consumer markets like Sweden and Switzerland where public awareness of sustainability issues is relatively high. One might expect that consumers in such contexts would be willing to pay a premium for environmentally responsible choices. Yet the data indicate otherwise—at least under current market conditions. This may reflect a range of explanations, including strategic pricing decisions by retailers, lack of consumer trust or understanding of climate labels, or limited willingness to pay in specific product categories.

These results raise a broader question: Are voluntary market mechanisms sufficient to drive climate-conscious consumption? Without stronger and more coordinated interventions, even well-designed climate labels may fail to influence consumer behavior meaningfully. Clearer policy support, more standardized and trusted labeling systems, and targeted consumer education may be necessary to bridge the gap between climate goals and real-world market dynamics.

## 5. Conclusion

This paper investigates whether climate-friendly food products are priced higher in consumer markets. Using detailed product-level data from supermarkets in Sweden and Switzerland, we examine the relationship between front-of-package climate impact scores and retail prices, controlling for product size, nutritional content, and fixed effects.

Our theoretical framework identifies two opposing sets of forces. On one side, eco-conscious consumer preferences, behavioral nudges from climate labels, and green product differentiation are expected to drive a positive price premium. On the other, market frictions such as psychological distance, competing priorities, and distrust in environmental labeling may suppress or reverse such effects. On the supply side, innovation and efficiency gains—consistent with the Porter Hypothesis—may enable producers to reduce emissions without raising prices, further complicating expectations of a green premium.

Contrary to intuitive expectations, we find no evidence that climate-friendly products command higher prices. In some product categories, better climate scores are actually associated with lower prices, indicating a negative premium—an outcome that presents a puzzle.

We argue that demand-side frictions—such as competing consumer priorities, psychological distance from climate issues, and skepticism toward labeling—likely suppress the emergence of price premiums intended to reward sustainable products. Our findings carry important implications for producers, retailers, and policymakers aiming to align climate objectives with effective incentive structures that support the transition to a more sustainable society.

Future research could employ experimental or survey-based methods to uncover the behavioral mechanisms underlying consumer responses—such as information fatigue, label skepticism, or price sensitivity. These insights are essential for designing more effective labeling schemes and policy tools that promote sustainable consumption. Looking ahead, it will also be important to explore how these dynamics may unfold in developing and transitioning economies, where climate-related labeling is not yet widely adopted. As these markets begin to consider sustainability labeling in the future, a deeper understanding of consumer awareness, trust, and preferences will be critical for guiding effective and context-sensitive implementation.

**Declaration of generative AI and AI-assisted technologies in the writing process**

During the preparation of this work the authors used GPT-4o in the writing process to improve the readability and language of the manuscript. After using this tool/service, the authors reviewed and edited the content as needed and take full responsibility for the content of the published article.

## References

Andersson, J. J. (2019). Carbon taxes and CO2 emissions: Sweden as a case study. *American Economic Journal: Economic Policy*, *11*(4), 1-30. https://doi.org/10.1257/pol.20170144

Conrad, K. (2005). Price competition and product differentiation when consumers care for the environment. *Environmental and Resource Economics*, *31*, 1-19. https://doi.org/10.1007/s10640-004-6977-8

Cranfield, J. A., & Magnusson, E. (2003). Canadian consumer's willingness-to-pay for pesticide free food products: an ordered probit analysis. *International Food and Agribusiness Management Review*, *6*(4).

Crippa, M., Solazzo, E., Guizzardi, Monforti-Ferrario, F., Tubiello, F.N., & Leip, A. (2021). Food systems are responsible for a third of global anthropogenic GHG emissions. *Nat Food* **2**, 198–209 https://doi.org/10.1038/s43016-021-00225-9

De-loyde, K., Pilling, M. A., Thornton, A., Spencer, G., & Maynard, O. M. (2025). Promoting sustainable diets using eco-labelling and social nudges: a randomised online experiment. *Behavioural Public Policy*, *9*(2), 426–442. https://doi.org/10.1017/bpp.2022.27

Donato, C. (2025). The Effect of Eco-Label Logos Visual Design on Consumers' Sustainability Perceptions: An Empirical Study. In: Eco-Label Visual Design and Sustainability. Palgrave Macmillan, Cham. https://doi.org/10.1007/978-3-031-82761-7_4

Feucht, Y., & Zander, K. (2018). Consumers' preferences for carbon labels and the underlying reasoning. A mixed methods approach in 6 European countries. *Journal of Cleaner Production*, *178*, 740-748. https://doi.org/10.1016/j.jclepro.2017.12.236

Finisterra do Paço, A. M., Barata Raposo, M. L., & Filho, W. L. (2009). Identifying the green consumer: A segmentation study. *Journal of Targeting, Measurement and Analysis for Marketing*, *17*, 17-25. https://doi.org/10.1057/jt.2008.28

Finger, R., Swinton, S. M., El Benni, N., & Walter, A. (2019). Precision farming at the nexus of agricultural production and the environment. *Annual Review of Resource Economics*, *11*(1), 313–335.

Getz, D., & Page, S. J. (2015). Progress and prospects for event tourism research. *Tourism Management*, *52*, 593–631. https://doi.org/10.1016/j.tourman.2015.03.007

Girod, B. (2012). Low-Carbon Society in Switzerland. In *Living in a Low-Carbon Society in 2050* (pp. 164-179). London: Palgrave Macmillan UK.

Gorton, M., Tocco, B., Yeh, C. H., & Hartmann, M. (2021). What determines consumers' use of eco-labels? Taking a close look at label trust. *Ecological Economics*, *189*, 107173. https://doi.org/10.1016/j.ecolecon.2021.107173

Hintermann, B., & Zarkovic, M. (2020). Carbon Pricing in Switzerland: A Fusion of Taxes, Command-and-Control, and Permit Markets. *DICE Report*, *18*(01), 35-41. https://hdl.handle.net/10419/225221

Hof, A. F., Den Elzen, M. G. J., & Mendoza Beltran, A. (2016). The EU 40% greenhouse gas emission reduction target by 2030 in perspective. *International Environmental Agreements: Politics, Law and Economics*, *16*, 375-392. https://doi.org/10.1007/s10784-016-9317-x

Hormio, S. (2023). Collective responsibility for climate change. *Wiley Interdisciplinary Reviews: Climate Change*, *14*(4), e830. https://doi.org/10.1002/wcc.830

Islam, M. Z., & Zheng, L. (2025). Why is it necessary to integrate circular economy practices for agri-food sustainability from a global perspective?. *Sustainable Development*, *33*(1), 600-620. https://doi.org/10.1002/sd.3135

Jäger-Waldau, A., Kougias, I., Taylor, N., & Thiel, C. (2020). How photovoltaics can contribute to GHG emission reductions of 55% in the EU by 2030. *Renewable and Sustainable Energy Reviews*, *126*, 109836. https://doi.org/10.1016/j.rser.2020.109836

Jaiswal, D., Kaushal, V., Singh, P. K., & Biswas, A. (2021). Green market segmentation and consumer profiling: a cluster approach to an emerging consumer market. *Benchmarking: An International Journal*, *28*(3), 792-812. https://doi.org/10.1108/BIJ-05-2020-0247

Karl, H., Ranné, O., & Macquarrie, J. (2019). The Spatial Dimension to Environmental Problems. In *Transition, Cohesion and Regional Policy in Central and Eastern Europe* (pp. 243-258). Routledge.

Lewandowsky, S., Ballard, T., Oberauer, K., & Benestad, R. (2016). A blind expert test of contrarian claims about climate data. *Global Environmental Change*, *39*, 91-97. https://doi.org/10.1016/j.gloenvcha.2016.04.013

Lombardi, G. V., Berni, R., & Rocchi, B. (2017). Environmental friendly food. Choice experiment to assess consumer's attitude toward "climate neutral" milk: the role of communication. *Journal of cleaner production*, *142*, 257-262. https://doi.org/10.1016/j.jclepro.2016.05.125

Mason, C. F. (2006). An economic model of ecolabeling. *Environmental Modeling & Assessment*, *11*, 131-143. https://doi.org/10.1007/s10666-005-9035-1

Moon, S. J., Costello, J. P., & Koo, D. M. (2016). The impact of consumer confusion from eco-labels on negative WOM, distrust, and dissatisfaction. *International Journal of Advertising*, *36*(2), 246–271. https://doi.org/10.1080/02650487.2016.1158223

Nakavachara, Voraprapa and Thongtai, Chanon and Chalidabhongse, Thanarat and Pharino, Chanathip, Ethical Appetite: Consumer Preferences and Price Premiums for Animal Welfare-Friendly Food Products (March 18, 2025). SSRN: https://ssrn.com/abstract=5249481

Olson, E. L. (2013). It's not easy being green: the effects of attribute tradeoffs on green product preference and choice. *Journal of the Academy of Marketing Science*, *41*, 171-184. https://doi.org/10.1007/s11747-012-0305-6

Porter, M. E., & Linde, C. V. D. (1995). Toward a new conception of the environment-competitiveness relationship. *Journal of economic perspectives*, *9*(4), 97-118. https://doi.org/10.1257/jep.9.4.97

Rondoni, A., & Grasso, S. (2021). Consumers behaviour towards carbon footprint labels on food: A review of the literature and discussion of industry implications. *Journal of Cleaner Production*, *301*, 127031. https://doi.org/10.1016/j.jclepro.2021.127031

Rubashkina, Y., Galeotti, M., & Verdolini, E. (2015). Environmental regulation and competitiveness: Empirical evidence on the Porter Hypothesis from European manufacturing sectors. *Energy policy*, *83*, 288-300. https://doi.org/10.1016/j.enpol.2015.02.014

Sandoff, A., & Schaad, G. (2009). Does EU ETS lead to emission reductions through trade? The case of the Swedish emissions trading sector participants. *Energy Policy*, *37*(10), 3967-3977.

Schmidt, J. R. (2008). Why Europe Leads on Climate Change. *Survival*, *50*(4), 83–96. https://doi.org/10.1080/00396330802328990

Scholz, K., Eriksson, M., & Strid, I. (2015). Carbon footprint of supermarket food waste. *Resources, Conservation and Recycling*, *94*, 56-65. https://doi.org/10.1016/j.resconrec.2014.11.016

Shaikh, S., Yamim, A. P., & Werle, C. O. (2024). Are all-encompassing better than one-trait sustainable labels? The influence of Eco-Score and organic labels on food perception and willingness to pay. *Appetite*, *203*, 107670. https://doi.org/10.1016/j.appet.2024.107670

Sinclair, M., Combet, E., Davis, T., & Papies, E. K. (2025). Sustainability in food-based dietary guidelines: a review of recommendations around meat and dairy consumption and their visual representation. *Annals of Medicine*, *57*(1). https://doi.org/10.1080/07853890.2025.2470252

Spence, A., & Pidgeon, N. (2009). Psychology, Climate Change & Sustainable Bahaviour. *Environment: Science and Policy for Sustainable Development*, *51*(6), 8–18. https://doi.org/10.1080/00139150903337217

Thaler, R. H., & Sunstein, C. R. (2021). *Nudge: The final edition*. Penguin.

United Nations Environment Programme (2024). *Food Waste Index Report 2024. Think Eat Save: Tracking Progress to Halve Global Food Waste*. https://wedocs.unep.org/20.500.11822/45230.

Watanabe, S. (2007). How animal psychology contributes to animal welfare. *Applied Animal Behaviour Science*, *106*(4), 193-202. https://doi.org/10.1016/j.applanim.2007.01.003

Woerdman, E. (2004). *The institutional economics of market-based climate policy* (Vol. 7). Elsevier. ISBN: 9780444515735.

Wölfl, A. and P. Sicari (2012), "Reducing Greenhouse Gas Emissions in a Cost Effective Way in Switzerland", *OECD Economics Department Working Papers*, No. 1002, OECD Publishing, Paris, https://doi.org/10.1787/5k8xff3tgd32-en.

Yogeesh, N. (2024). Climate Change Adaptation Challenges Require Technical Expertise and Data Management. In *Creating Pathways for Prosperity* (pp. 45-65). Emerald Publishing Limited. https://doi.org/10.1108/978-1-83549-121-820241005

Figure 1: Research Framework

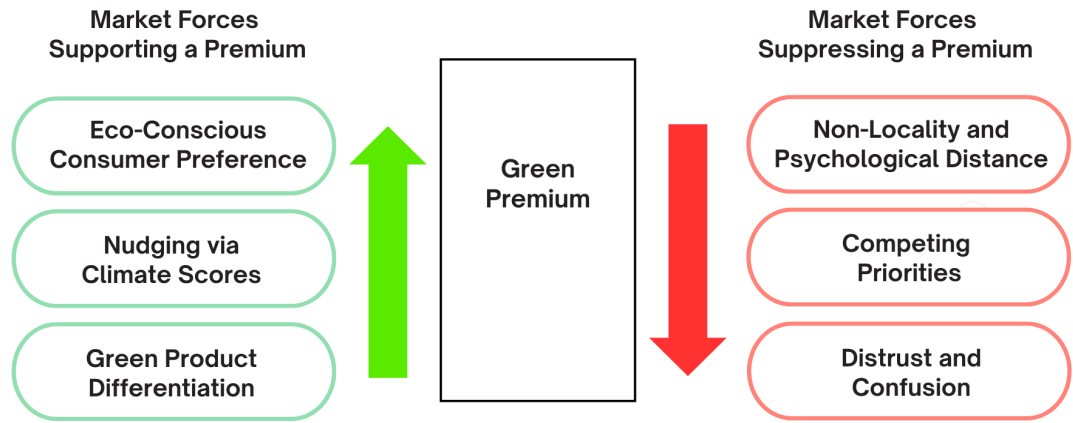

Table 1: Summary Statistics of Food Products from Swedish Supermarket

| Variable | Obs | Mean | Std. dev. | Min | Max |
|---|---|---|---|---|---|
| LnPrice | 5,458 | 3.48 | 0.64 | 0.92 | 7.60 |
| PackageSize | 5,458 | 357.93 | 328.67 | 0.22 | 4,500.00 |
| Climate | 2,470 | 3.89 | 0.86 | 1.00 | 5.00 |
| Carb | 5,398 | 101.47 | 171.85 | 0.00 | 3,510.00 |
| Fat | 5,414 | 39.80 | 61.04 | 0.00 | 800.00 |
| Protein | 5,404 | 31.80 | 45.99 | 0.00 | 638.00 |
| Salt | 5,344 | 4.57 | 39.79 | 0.00 | 2,000.00 |
| Energy | 5,436 | 911.20 | 988.42 | 0.00 | 15,525.00 |

Table 2: Food Products by Category from Swedish Supermarket

| Category Level 1 | Freq. | Percent | Cum. |
|---|---|---|---|
| Bread & Bakery | 626 | 11.47 | 11.47 |
| Candy, Ice Cream & Snacks | 754 | 13.81 | 25.28 |
| Cheese | 403 | 7.38 | 32.67 |
| Dairy & Eggs | 420 | 7.7 | 40.36 |
| Delicacies | 93 | 1.7 | 42.07 |
| Freezer | 431 | 7.9 | 49.96 |
| Larder | 1,457 | 26.69 | 76.66 |
| Meat, Poultry & Charcuterie | 599 | 10.97 | 87.63 |
| Ready meals & snacks | 307 | 5.62 | 93.26 |
| Spices & Seasonings | 368 | 6.74 | 100 |
| Total | 5,458 | 100 | |

Table 3: Regression Results from Swedish Supermarket
(All vs. Europe vs. Non-Europe)

| VARIABLES | (1) LnPrice | (2) LnPrice | (3) LnPrice |
|---|---|---|---|
| Climate | -0.0730 | -0.0868* | 0.00827 |
|  | (0.0471) | (0.0499) | (0.0473) |
| Package Size | 0.000393*** | 0.000379*** | 0.00214 |
|  | (0.000120) | (0.000119) | (0.00216) |
| Carb | -6.25e-06 | 4.70e-05 | -0.0159** |
|  | (0.000261) | (0.000272) | (0.00609) |
| Fat | 0.00128** | 0.00135** | -0.0321* |
|  | (0.000566) | (0.000557) | (0.0184) |
| Protein | 0.00182** | 0.00188** | -0.0104 |
|  | (0.000782) | (0.000786) | (0.0119) |
| Salt | 0.00201 | 0.00188 | -0.0104 |
|  | (0.00130) | (0.00126) | (0.0719) |
| Energy | -9.64e-06 | -1.26e-05 | 0.00333* |
|  | (2.45e-05) | (2.23e-05) | (0.00167) |
| Constant | 3.929*** | 3.992*** | 3.455*** |
|  | (0.265) | (0.276) | (0.414) |
|  |  |  |  |
| Observations | 2,314 | 2,177 | 137 |
| R-squared | 0.729 | 0.731 | 0.886 |
| Country Dummy | Yes | Yes | Yes |
| Category Dummy | CategoryL3 | CategoryL3 | CategoryL3 |
| Cluster | CategoryL3 | CategoryL3 | CategoryL3 |
| Group | All | Europe | Non-Europe |
| (All vs. Europe vs. non-Europe) |  |  |  |

Robust standard errors in parentheses
*** $p<0.01$, ** $p<0.05$, * $p<0.1$

Table 4: Regression Results from Swedish Supermarket

(All vs. by Category Level 1)

| VARIABLES | (1) LnPrice | (2) LnPrice | (3) LnPrice | (4) LnPrice | (5) LnPrice | (6) LnPrice | (7) LnPrice | (8) LnPrice | (9) LnPrice | (10) LnPrice | (11) LnPrice |
|---|---|---|---|---|---|---|---|---|---|---|---|
| Climate | -0.0730 | -0.0108 | -0.0571 | -0.135 | 0.0104 | 1.834 | -0.116* | -0.0385 | -0.221** | -0.0530** | 0.0366 |
|  | (0.0471) | (0.0435) | (0.126) | (0.239) | (0.0918) | (1.106) | (0.0576) | (0.0660) | (0.105) | (0.0211) | (0.0353) |
| Package Size | 0.000393*** | 0.00169* | 0.00985** | 0.000371 | 7.32e-05 | 0.000670 | 0.000218 | -2.92e-06 | -0.000653 | -0.000584 | 0.000466 |
|  | (0.000120) | (0.000860) | (0.00400) | (0.00117) | (0.000230) | (0.000779) | (0.000235) | (0.000178) | (0.000745) | (0.000467) | (0.00169) |
| Carb | -6.25e-06 | -0.00195* | -0.00381 | -0.00790 | -0.00211 | -0.00959 | 0.00356 | -0.000960 | 0.00308 | -0.00654 | -0.00107 |
|  | (0.000261) | (0.00100) | (0.00322) | (0.00785) | (0.00801) | (0.0354) | (0.00273) | (0.00136) | (0.00420) | (0.00550) | (0.00823) |
| Fat | 0.00128** | 0.00330*** | 0.00596 | -0.0204 | -0.00855 | -0.00462 | 0.00538 | 3.12e-05 | 0.00102 | -0.0148 | -0.00776 |
|  | (0.000566) | (0.000993) | (0.00789) | (0.0177) | (0.0174) | (0.00391) | (0.00580) | (0.00296) | (0.00159) | (0.0117) | (0.0178) |
| Protein | 0.00182** | 0.000171 | -0.0105 | -0.00905 | 0.00709 | 0.0971 | 0.00137 | -0.000741 | 0.00750** | -0.00232 | 0.0191*** |
|  | (0.000782) | (0.00425) | (0.00674) | (0.00704) | (0.00807) | (0.0551) | (0.00207) | (0.00193) | (0.00349) | (0.00624) | (0.00635) |
| Salt | 0.00201 | 0.0216 | -0.0294 | -0.00392 | -0.0259 | 0.0453** | 0.0142 | 0.0111** | 0.00996 | 0.0566 | 0.00137 |
|  | (0.00130) | (0.0137) | (0.0304) | (0.0259) | (0.0168) | (0.0124) | (0.0174) | (0.00456) | (0.0110) | (0.0405) | (0.00167) |
| Energy | -9.64e-06 | -2.22e-05 | -0.00113 | 0.00251 | 0.00120 | 5.05e-05 | -0.000401 | 0.000310 | -6.85e-05 | 0.00216 | 0.000978 |
|  | (2.45e-05) | (1.41e-05) | (0.00121) | (0.00207) | (0.00199) | (7.49e-05) | (0.000605) | (0.000342) | (0.000157) | (0.00138) | (0.00196) |
| Constant | 3.929*** | 2.437*** | 3.297*** | 3.715*** | 3.139*** | -3.943 | 3.114*** | 3.474*** | 4.417*** | 3.738*** | 2.296*** |
|  | (0.265) | (0.174) | (0.393) | (0.910) | (0.418) | (5.056) | (0.323) | (0.354) | (0.412) | (0.377) | (0.519) |
| Observations | 2,314 | 261 | 209 | 213 | 230 | 59 | 160 | 548 | 360 | 112 | 162 |
| R-squared | 0.729 | 0.633 | 0.830 | 0.762 | 0.852 | 0.836 | 0.830 | 0.656 | 0.650 | 0.892 | 0.820 |
| Country Dummy | Yes | Yes | Yes | Yes | Yes | Yes | Yes | Yes | Yes | Yes | Yes |
| Category Dummy | CategoryL3 | CategoryL3 | CategoryL3 | CategoryL3 | CategoryL3 | CategoryL3 | CategoryL3 | CategoryL3 | CategoryL3 | CategoryL3 | CategoryL3 |
| Cluster | CategoryL3 | CategoryL3 | CategoryL3 | CategoryL3 | CategoryL3 | CategoryL3 | CategoryL3 | CategoryL3 | CategoryL3 | CategoryL3 | CategoryL3 |
| Group (All vs. CategoryL1) | All | Bread & Bakery | Candy & Ice Cream & Snacks | Cheese | Dairy & Eggs | Delicacies | Freezer | Larder | Meat & Poultry & Charcuterie | Ready meals & snacks | Spices & Seasonings |

Robust standard errors in parentheses

*** p<0.01, ** p<0.05, * p<0.1

Table 5: Regression Results from Swiss Supermarket
(All vs. Europe vs. Non-Europe)

| VARIABLES | (1) LnPrice | (2) LnPrice | (3) LnPrice |
|---|---|---|---|
| Climate | -0.0549*** | -0.0553*** | 0.000639 |
|  | (0.0186) | (0.0200) | (0.0924) |
| Package Size | 0.000270*** | 0.000250*** | -8.74e-05 |
|  | (8.51e-05) | (8.07e-05) | (0.000354) |
| Carb | 0.00161** | 0.000787 | -0.00549 |
|  | (0.000622) | (0.00206) | (0.00611) |
| Fat | 0.00602*** | 0.00475 | -0.0145 |
|  | (0.00131) | (0.00451) | (0.0148) |
| Protein | 0.00443*** | 0.00363 | -0.00331 |
|  | (0.00124) | (0.00258) | (0.00741) |
| Salt | -0.000997 | -0.000147 | 0.0370** |
|  | (0.00115) | (0.00169) | (0.0180) |
| Energy | -0.000479*** | -0.000275 | 0.00141 |
|  | (0.000149) | (0.000514) | (0.00159) |
| Constant | 1.609*** | 1.548*** | 2.271*** |
|  | (0.199) | (0.203) | (0.361) |
|  |  |  |  |
| Observations | 2,810 | 2,599 | 211 |
| R-squared | 0.680 | 0.687 | 0.771 |
| Country Dummy | Yes | Yes | Yes |
| Category Dummy | CategoryL3 | CategoryL3 | CategoryL3 |
| Cluster | CategoryL3 | CategoryL3 | CategoryL3 |
| Group | All | Europe | non-Europe |
| (All vs. Europe vs. non-Europe) |  |  |  |

Robust standard errors in parentheses
*** p<0.01, ** p<0.05, * p<0.1

Table 6: Regression Results from Swiss Supermarket
(All vs. by Category Level 1)

| VARIABLES | (1) LnPrice | (2) LnPrice | (3) LnPrice | (4) LnPrice | (5) LnPrice | (6) LnPrice | (7) LnPrice | (8) LnPrice |
|---|---|---|---|---|---|---|---|---|
| Climate | -0.0549*** | -0.0903* | -0.0706 | -0.0938* | -0.265 | -0.0589** | -0.0670** | 0.0454 |
|  | (0.0186) | (0.0481) | (0.0543) | (0.0486) | (0.322) | (0.0270) | (0.0320) | (0.0454) |
| Package Size | 0.000270*** | 0.000261 | 0.000304 | -5.79e-06 | 0.000374* | -0.00102* | 0.000488*** | 0.000748 |
|  | (8.51e-05) | (0.000737) | (0.000218) | (0.000132) | (0.000193) | (0.000580) | (0.000100) | (0.000490) |
| Carb | 0.00161** | 0.00177 | -0.000646 | -0.00174 | 0.00530 | -0.000947 | 0.00636* | -0.00678 |
|  | (0.000622) | (0.00534) | (0.00169) | (0.00952) | (0.0176) | (0.0159) | (0.00340) | (0.00496) |
| Fat | 0.00602*** | 0.00548 | 0.00373 | -0.00964 | 0.00558 | -0.00610 | 0.0171** | -0.0121 |
|  | (0.00131) | (0.0120) | (0.00317) | (0.0196) | (0.0427) | (0.0326) | (0.00782) | (0.0106) |
| Protein | 0.00443*** | 0.00158 | 0.00559** | 2.26e-06 | 0.000989 | 0.00887 | 0.0114*** | -0.0138** |
|  | (0.00124) | (0.00606) | (0.00202) | (0.00913) | (0.0169) | (0.0152) | (0.00393) | (0.00655) |
| Salt | -0.000997 | 0.0112 | 0.00511 | 0.0229*** | -0.0497*** | 0.0385*** | -0.00150 | -0.0189* |
|  | (0.00115) | (0.0106) | (0.0175) | (0.00714) | (0.0153) | (0.0109) | (0.000964) | (0.0105) |
| Energy | -0.000479*** | -0.000448 | -0.000203 | 0.000859 | -0.00148 | 0.000498 | -0.00185** | 0.00174 |
|  | (0.000149) | (0.00137) | (0.000380) | (0.00221) | (0.00447) | (0.00363) | (0.000847) | (0.00120) |
| Constant | 1.609*** | 1.384*** | 0.759** | 1.776*** | 2.719 | 0.885*** | 1.679*** | 0.939*** |
|  | (0.199) | (0.216) | (0.299) | (0.212) | (1.858) | (0.281) | (0.187) | (0.245) |
| Observations | 2,810 | 344 | 532 | 272 | 112 | 376 | 583 | 591 |
| R-squared | 0.680 | 0.488 | 0.754 | 0.718 | 0.555 | 0.744 | 0.428 | 0.533 |
| Country Dummy | Yes | Yes | Yes | Yes | Yes | Yes | Yes | Yes |
| Category Dummy | CategoryL3 | CategoryL3 | CategoryL3 | CategoryL3 | CategoryL3 | CategoryL3 | CategoryL3 | CategoryL3 |
| Cluster | CategoryL3 | CategoryL3 | CategoryL3 | CategoryL3 | CategoryL3 | CategoryL3 | CategoryL3 | CategoryL3 |
| Group (All vs. CategoryL1) | All | Bread & pastries & breakfast | Dairy & eggs & fresh convenience food | Frozen food | Fruits & vegetables | Meat & fish | Pasta & condiments & canned food | Snacks & sweets |

Robust standard errors in parentheses
*** p<0.01, ** p<0.05, * p<0.1

Appendix 1: Details on Categories of Food Products from Swedish Supermarket (Page 1/2)

| CategoryLevel1 | CategoryLevel2 | CategoryLevel3 | _freq |
|---|---|---|---|
| Bread & Bakery | Bake off & bake at home | Bread | 9 |
| Bread & Bakery | Bakery | Baguettes & Buns | 5 |
| Bread & Bakery | Bakery | Bread with fruit and nuts | 1 |
| Bread & Bakery | Bakery | Country bread | 1 |
| Bread & Bakery | Bakery | Pastries | 12 |
| Bread & Bakery | Cheese Biscuits, Corn Cakes & Snacks | Cheese Biscuits & Snacks | 52 |
| Bread & Bakery | Cheese Biscuits, Corn Cakes & Snacks | Corn Cakes & Rice Cakes | 34 |
| Bread & Bakery | Cheese Biscuits, Corn Cakes & Snacks | Grissini & Krustader | 5 |
| Bread & Bakery | Cheese Biscuits, Corn Cakes & Snacks | Rusks | 13 |
| Bread & Bakery | Crispbread | Crispbread portion | 65 |
| Bread & Bakery | Crispbread | Crispbread whole | 19 |
| Bread & Bakery | Crispbread | Gluten-free crispbread | 16 |
| Bread & Bakery | Crispbread | Hard flatbread | 8 |
| Bread & Bakery | Dark bread | Gluten-free dark bread | 12 |
| Bread & Bakery | Dark bread | Sourdough bread | 2 |
| Bread & Bakery | Dark bread | Wholemeal bread | 43 |
| Bread & Bakery | Light bread | Country bread | 7 |
| Bread & Bakery | Light bread | Flatbread & Wheat Cookies | 20 |
| Bread & Bakery | Light bread | Gluten-free white bread | 14 |
| Bread & Bakery | Light bread | Sourdough bread | 7 |
| Bread & Bakery | Light bread | Toast | 13 |
| Bread & Bakery | Pastries | Buns & Wheat Length | 11 |
| Bread & Bakery | Pastries | Cake base & Meringue | 6 |
| Bread & Bakery | Pastries | Cookies & Biscuits | 128 |
| Bread & Bakery | Pastries | Gluten-free pastries | 23 |
| Bread & Bakery | Pastries | Soft cakes & Sweets | 57 |
| Bread & Bakery | Sausage Bread, Hamburger Bread & Other | Gluten-free sausage bread & Other | 4 |
| Bread & Bakery | Sausage Bread, Hamburger Bread & Other | Hamburger bread | 7 |
| Bread & Bakery | Sausage Bread, Hamburger Bread & Other | Meal bread & street food | 13 |
| Bread & Bakery | Sausage Bread, Hamburger Bread & Other | Sausage bread | 19 |
| Bread & Bakery | Sausage Bread, Hamburger Bread & Other | Tortilla Bread & Taco Bread | 5 |
| Candy, Ice Cream & Snacks | Candy | Candy bags | 165 |
| Candy, Ice Cream & Snacks | Candy | Licorice | 18 |
| Candy, Ice Cream & Snacks | Candy | Tablet cases & Pieces | 30 |
| Candy, Ice Cream & Snacks | Chocolate | Chocolate Pieces | 90 |
| Candy, Ice Cream & Snacks | Chocolate | Dark chocolate bars | 32 |
| Candy, Ice Cream & Snacks | Chocolate | Gift boxes | 33 |
| Candy, Ice Cream & Snacks | Chocolate | Light chocolate bars | 64 |
| Candy, Ice Cream & Snacks | Crisp | Crisps | 84 |
| Candy, Ice Cream & Snacks | Crisp | Dip mix | 12 |
| Candy, Ice Cream & Snacks | Crisp | Lens chips | 4 |
| Candy, Ice Cream & Snacks | Crisp | Other chips | 13 |
| Candy, Ice Cream & Snacks | Crisp | Tortilla chips | 11 |
| Candy, Ice Cream & Snacks | Ice cream | Lactose-free ice cream | 1 |
| Candy, Ice Cream & Snacks | Ice cream | Sprinkles & Ice Cream Sauce | 12 |
| Candy, Ice Cream & Snacks | Ice cream | Waffle cones | 3 |
| Candy, Ice Cream & Snacks | Nuts & Dried Fruits | Cashews | 13 |
| Candy, Ice Cream & Snacks | Nuts & Dried Fruits | Dried fruit | 18 |
| Candy, Ice Cream & Snacks | Nuts & Dried Fruits | Other nuts | 24 |
| Candy, Ice Cream & Snacks | Nuts & Dried Fruits | Peanuts | 12 |
| Candy, Ice Cream & Snacks | Snacks | Beer Sausages & Snacks | 2 |
| Candy, Ice Cream & Snacks | Snacks | Cheese bows | 15 |
| Candy, Ice Cream & Snacks | Snacks | Corn snacks | 1 |
| Candy, Ice Cream & Snacks | Snacks | Other snacks | 11 |
| Candy, Ice Cream & Snacks | Snacks | Popcorn | 11 |
| Candy, Ice Cream & Snacks | Snacks | Salty sticks | 1 |
| Candy, Ice Cream & Snacks | Throat Lozenges & Chewing Gum | Chewing gum | 43 |
| Candy, Ice Cream & Snacks | Throat Lozenges & Chewing Gum | Throat lozenges | 31 |
| Cheese | Cold cuts | Cheddar | 10 |
| Cheese | Cold cuts | Cream cheese | 46 |
| Cheese | Cold cuts | Dairy-free spread cheese | 3 |
| Cheese | Cold cuts | Edam | 5 |
| Cheese | Cold cuts | Gouda | 11 |
| Cheese | Cold cuts | Grevé | 11 |
| Cheese | Cold cuts | Hamburger cheese | 6 |
| Cheese | Cold cuts | Household cheese | 10 |
| Cheese | Cold cuts | Manor | 15 |
| Cheese | Cold cuts | Other cheeses | 34 |
| Cheese | Cold cuts | Port Salut | 5 |
| Cheese | Cold cuts | Priest's cheese | 15 |
| Cheese | Cold cuts | Soft cheese | 42 |
| Cheese | Cold cuts | Västerbotten cheese | 4 |
| Cheese | Cold cuts | Whey butter & Whey cheese | 8 |
| Cheese | Cooking cheese | Cream cheese | 4 |
| Cheese | Cooking cheese | Feta Cheese & Salad Cheese | 34 |
| Cheese | Cooking cheese | Grated cheese | 36 |
| Cheese | Cooking cheese | Halloumi & Grilling Cheese | 15 |
| Cheese | Cooking cheese | Mozzarella | 18 |
| Cheese | Cooking cheese | Parmesan & other hard cheeses | 18 |
| Cheese | Dessert cheese | Blue cheese | 16 |
| Cheese | Dessert cheese | Brie | 14 |
| Cheese | Dessert cheese | Camembert | 3 |
| Cheese | Dessert cheese | Chèvre | 8 |
| Cheese | Dessert cheese | Gorgonzola | 1 |
| Cheese | Dessert cheese | Gruyère | 2 |
| Cheese | Dessert cheese | Manchego | 2 |
| Cheese | Dessert cheese | Other dessert cheese | 6 |
| Cheese | Dessert cheese | St Agur | 1 |

| CategoryLevel1 | CategoryLevel2 | CategoryLevel3 | _freq |
|---|---|---|---|
| Dairy & Eggs | Butter & Margarine | Dairy-free butter & margarine | 1 |
| Dairy & Eggs | Butter & Margarine | Food & Baking Butter | 15 |
| Dairy & Eggs | Butter & Margarine | Lactose-free butter & margarine | 3 |
| Dairy & Eggs | Butter & Margarine | Other cooking fat | 1 |
| Dairy & Eggs | Butter & Margarine | Tabletop butter & margarine | 46 |
| Dairy & Eggs | Chilled snacks & desserts | Cheesecake | 3 |
| Dairy & Eggs | Chilled snacks & desserts | Chilled desserts | 10 |
| Dairy & Eggs | Chilled snacks & desserts | Chilled small meals | 22 |
| Dairy & Eggs | Chilled snacks & desserts | Custard | 1 |
| Dairy & Eggs | Chilled snacks & desserts | Drink small meal | 2 |
| Dairy & Eggs | Chilled snacks & desserts | Plant-based snacks & desserts | 4 |
| Dairy & Eggs | Chilled snacks & desserts | Protein targets | 21 |
| Dairy & Eggs | Cooking dairies | Crème fraiche flavored | 2 |
| Dairy & Eggs | Cooking dairies | Crème fraiche natural | 1 |
| Dairy & Eggs | Cooking dairies | Plant-based crème fraiche & sour cream | 9 |
| Dairy & Eggs | Cream | Whipped cream | 1 |
| Dairy & Eggs | Eggs & Yeast | Egg | 7 |
| Dairy & Eggs | Eggs & Yeast | Yeast | 4 |
| Dairy & Eggs | Quark & Cottage Cheese | Flavored cottage cheese | 2 |
| Dairy & Eggs | Quark & Cottage Cheese | Flavoured quark | 25 |
| Dairy & Eggs | Quark & Cottage Cheese | Natural cottage cheese | 12 |
| Dairy & Eggs | Quark & Cottage Cheese | Natural quark | 4 |
| Dairy & Eggs | Yogurt & Sour Cream | Cooking yogurt | 13 |
| Dairy & Eggs | Yogurt & Sour Cream | Drinking yogurt | 9 |
| Dairy & Eggs | Yogurt & Sour Cream | Flavored file | 17 |
| Dairy & Eggs | Yogurt & Sour Cream | Flavored yogurt | 68 |
| Dairy & Eggs | Yogurt & Sour Cream | Health Yogurt & Health File | 10 |
| Dairy & Eggs | Yogurt & Sour Cream | Lactose-free yogurt & Sour cream | 49 |
| Dairy & Eggs | Yogurt & Sour Cream | Multipack yogurt | 11 |
| Dairy & Eggs | Yogurt & Sour Cream | Natural file | 19 |
| Dairy & Eggs | Yogurt & Sour Cream | Natural yogurt | 16 |
| Dairy & Eggs | Yogurt & Sour Cream | Plant-based yogurt | 12 |
| Delicacies | Cheese platter | Accessories for the cheese platter | 42 |
| Delicacies | Cheese platter | Biscuits for the cheese platter | 2 |
| Delicacies | Cheese platter | Hard dessert cheeses | 1 |
| Delicacies | Cheese platter | Soft dessert cheeses | 9 |
| Delicacies | Delicatessen | Other deli products | 15 |
| Delicacies | Delicatessen | Salami | 14 |
| Delicacies | Delicatessen | Smoked fish | 3 |
| Delicacies | Pastries & desserts | Confectionery | 7 |
| Freezer | Berries & Fruits | Blueberry | 4 |
| Freezer | Berries & Fruits | Fruit | 6 |
| Freezer | Berries & Fruits | Other berries | 9 |
| Freezer | Berries & Fruits | Raspberry | 4 |
| Freezer | Berries & Fruits | Strawberries | 5 |
| Freezer | Bird | Fillet | 14 |
| Freezer | Bird | Parts | 12 |
| Freezer | Bird | Prepared | 21 |
| Freezer | Bird | Whole | 2 |
| Freezer | Bread & Dessert | Bread | 7 |
| Freezer | Bread & Dessert | Cake, Pie & Dessert | 19 |
| Freezer | Bread & Dessert | Cooking & Baking | 3 |
| Freezer | Bread & Dessert | Gluten-free | 11 |
| Freezer | Bread & Dessert | Pastries | 5 |
| Freezer | Fish & Seafood | Breaded fish | 13 |
| Freezer | Fish & Seafood | Frozen cod | 9 |
| Freezer | Fish & Seafood | Frozen salmon | 5 |
| Freezer | Fish & Seafood | Other frozen fish | 7 |
| Freezer | Fish & Seafood | Other seafood | 4 |
| Freezer | Fish & Seafood | Shrimp | 12 |
| Freezer | Greens | Greens | 45 |
| Freezer | Greens | Potato products | 4 |
| Freezer | Greens | Stir-fry vegetables | 4 |
| Freezer | Greens | Vegetable mixes | 7 |
| Freezer | Ice cream | Ice cream package | 3 |
| Freezer | Ice cream | Multipack of Ice Cream | 8 |
| Freezer | Meat & Game | Game | 4 |
| Freezer | Meat & Game | Meat | 2 |
| Freezer | Meat & Game | Meatballs & Burgers | 18 |
| Freezer | Meat & Game | Other meat & game | 5 |
| Freezer | Portion dishes | Chicken | 19 |
| Freezer | Portion dishes | Fish | 4 |
| Freezer | Portion dishes | Meat | 32 |
| Freezer | Portion dishes | Vegetarian | 10 |
| Freezer | Ready to heat | Other ready-to-eat food | 24 |
| Freezer | Ready to heat | Pasties | 4 |
| Freezer | Ready to heat | Pie | 1 |
| Freezer | Ready to heat | Pizza | 61 |
| Freezer | Ready to heat | Soup | 1 |
| Freezer | Vegetarian | Mince, Fillet & Pieces | 1 |
| Freezer | Vegetarian | Ready to heat | 2 |

| CategoryLevel1 | CategoryLevel2 | CategoryLevel3 | _freq |
|---|---|---|---|
| Larder | Asian foods | Accessory | 15 |
| Larder | Asian foods | Bread | 2 |
| Larder | Asian foods | Canned vegetables | 6 |
| Larder | Asian foods | Coconut products | 1 |
| Larder | Asian foods | Rice & Noodles | 27 |
| Larder | Asian foods | Sauce | 18 |
| Larder | Asian foods | Seasoning | 11 |
| Larder | Asian foods | Soup | 3 |
| Larder | Asian foods | Spice mix | 15 |
| Larder | Asian foods | Spices | 2 |
| Larder | Baking | Baking Powder & Vanilla Sugar | 5 |
| Larder | Baking | Baking mix | 17 |
| Larder | Baking | Cocoa & Block Chocolate | 12 |
| Larder | Baking | Decoration | 24 |
| Larder | Baking | Flour | 60 |
| Larder | Baking | Honey & Syrup | 24 |
| Larder | Baking | Other things to bake | 24 |
| Larder | Baking | Sugar & Sweetening | 31 |
| Larder | Beans & Lentils | Beans | 23 |
| Larder | Beans & Lentils | Lenses | 12 |
| Larder | Beans & Lentils | Peas | 8 |
| Larder | Beans & Lentils | Quinoa | 3 |
| Larder | Canned meat and fish | Fish balls | 7 |
| Larder | Canned meat and fish | Mackerel & Anchovies | 9 |
| Larder | Canned meat and fish | Meat | 16 |
| Larder | Canned meat and fish | Shellfish | 3 |
| Larder | Canned meat and fish | Tuna | 5 |
| Larder | Canned vegetables | Beans | 8 |
| Larder | Canned vegetables | Beetroot | 10 |
| Larder | Canned vegetables | Corn | 7 |
| Larder | Canned vegetables | Cucumber | 23 |
| Larder | Canned vegetables | Mushroom | 7 |
| Larder | Canned vegetables | Oliver | 16 |
| Larder | Canned vegetables | Other canned vegetables | 21 |
| Larder | Canned vegetables | Pickles & Marinated | 14 |
| Larder | Canned vegetables | Tomato | 43 |
| Larder | Cereal & Muesli | Breakfast cereal | 38 |
| Larder | Cereal & Muesli | Grain | 40 |
| Larder | Cereal & Muesli | Granola & Crunch | 53 |
| Larder | Cereal & Muesli | Muesli | 34 |
| Larder | Cereal & Muesli | Pillows, Rings & Pouffs | 27 |
| Larder | Desserts | Bars | 47 |
| Larder | Desserts | Chocolate Pudding & Mousse | 6 |
| Larder | Desserts | Custard & dessert sauce | 5 |
| Larder | Desserts | Dried fruit | 49 |
| Larder | Desserts | Fruit Cream & Soups | 1 |
| Larder | Desserts | Fruit preserves | 19 |
| Larder | Jams, marmalades & other | Applesauce & Fruit Puree | 7 |
| Larder | Jams, marmalades & other | Jam | 54 |
| Larder | Jams, marmalades & other | Jelly | 5 |
| Larder | Jams, marmalades & other | Marmalade | 46 |
| Larder | Jams, marmalades & other | Peanut butter & hazelnut cream | 24 |
| Larder | Nuts & Dried Fruits | Kernels & Seeds | 17 |
| Larder | Nuts & Dried Fruits | Other nuts | 21 |
| Larder | Nuts & Dried Fruits | Walnuts | 4 |
| Larder | Pasta & Pasta Sauce | Fresh pasta | 17 |
| Larder | Pasta & Pasta Sauce | Gluten-free pasta | 17 |
| Larder | Pasta & Pasta Sauce | Lasagna | 7 |
| Larder | Pasta & Pasta Sauce | Macaroni | 12 |
| Larder | Pasta & Pasta Sauce | Noodles | 8 |
| Larder | Pasta & Pasta Sauce | Pasta sauce | 30 |
| Larder | Pasta & Pasta Sauce | Pesto & Tapenade | 25 |
| Larder | Pasta & Pasta Sauce | Shape paste | 79 |
| Larder | Pasta & Pasta Sauce | Spaghetti | 24 |
| Larder | Rice, mash & grains | Bulgur, Couscous & Wheat | 12 |
| Larder | Rice, mash & grains | Mashed Potatoes & Root Mashed | 6 |
| Larder | Rice, mash & grains | Rice | 54 |
| Larder | Tex Mex | Dip & Accessories | 11 |
| Larder | Tex Mex | Nachos | 17 |
| Larder | Tex Mex | Other Tex Mex | 6 |
| Larder | Tex Mex | Sauce & Salsa | 35 |
| Larder | Tex Mex | Spice mix | 26 |
| Larder | Tex Mex | Taco Shell | 6 |
| Larder | Tex Mex | Tortilla bread | 20 |
| Larder | World Food | Dessert | 2 |
| Larder | World Food | Food | 1 |
| Larder | World Food | Larder | 1 |
| Larder | World Food | Preserves | 6 |
| Larder | World Food | Seasoning | 6 |
| Meat, Poultry & Charcuterie | Beef | Bit | 14 |
| Meat, Poultry & Charcuterie | Beef | Discs | 23 |
| Meat, Poultry & Charcuterie | Beef | Mince & Burgers | 9 |
| Meat, Poultry & Charcuterie | Beef | Shredded & Diced | 6 |
| Meat, Poultry & Charcuterie | Calf & Game | Calf | 4 |
| Meat, Poultry & Charcuterie | Chark | Bacon & Roast Pork | 29 |
| Meat, Poultry & Charcuterie | Chark | Black pudding | 5 |
| Meat, Poultry & Charcuterie | Chark | Cold Smoked & Air Dried | 21 |
| Meat, Poultry & Charcuterie | Chark | Kassler & Ham | 9 |
| Meat, Poultry & Charcuterie | Chark | Meatballs & Minced Meat Products | 17 |
| Meat, Poultry & Charcuterie | Chicken & Other poultry | Chicken parts | 21 |
| Meat, Poultry & Charcuterie | Chicken & Other poultry | Other birds | 5 |
| Meat, Poultry & Charcuterie | Chicken & Other poultry | Whole Chicken & Fillet | 27 |
| Meat, Poultry & Charcuterie | Cold cuts & Deli | Cooked ham | 6 |
| Meat, Poultry & Charcuterie | Cold cuts & Deli | Jam & Pâté | 8 |
| Meat, Poultry & Charcuterie | Cold cuts & Deli | Other cold cuts | 28 |
| Meat, Poultry & Charcuterie | Cold cuts & Deli | Pâté | 26 |
| Meat, Poultry & Charcuterie | Cold cuts & Deli | Sausage | 98 |
| Meat, Poultry & Charcuterie | Cold cuts & Deli | Smoked ham | 31 |
| Meat, Poultry & Charcuterie | Cold cuts & Deli | Toppings of poultry | 17 |
| Meat, Poultry & Charcuterie | Lamb | Bit | 9 |
| Meat, Poultry & Charcuterie | Lamb | Discs | 1 |
| Meat, Poultry & Charcuterie | Lamb | Mince & Burgers | 1 |
| Meat, Poultry & Charcuterie | Pork | Bit | 28 |
| Meat, Poultry & Charcuterie | Pork | Discs | 13 |
| Meat, Poultry & Charcuterie | Pork | Mince | 3 |
| Meat, Poultry & Charcuterie | Pork | Seasoned & Marinated | 1 |
| Meat, Poultry & Charcuterie | Pork | Shredded & Diced | 2 |
| Meat, Poultry & Charcuterie | Sausage | Barbecue sausages | 24 |
| Meat, Poultry & Charcuterie | Sausage | Beer sausage | 16 |
| Meat, Poultry & Charcuterie | Sausage | Bratwurst | 2 |
| Meat, Poultry & Charcuterie | Sausage | Chorizo | 14 |
| Meat, Poultry & Charcuterie | Sausage | Falukorv | 12 |
| Meat, Poultry & Charcuterie | Sausage | Hot & Viennese sausages | 17 |
| Meat, Poultry & Charcuterie | Sausage | Lamb sausage | 1 |
| Meat, Poultry & Charcuterie | Sausage | Other sausages | 43 |
| Meat, Poultry & Charcuterie | Sausage | Prince's sausage | 1 |
| Meat, Poultry & Charcuterie | Sausage | Salsiccia | 7 |
| Ready meals & snacks | Mayonnaise salads | Other Stir-Fries, Salads | 29 |
| Ready meals & snacks | Mayonnaise salads | Potato salad | 15 |
| Ready meals & snacks | Mayonnaise salads | Shrimp & Seafood Salad | 13 |
| Ready meals & snacks | Potato products | French Fries & Strips | 16 |
| Ready meals & snacks | Potato products | Other potato products | 2 |
| Ready meals & snacks | Potato products | Potato Buns & Rösti | 5 |
| Ready meals & snacks | Potato products | Potato gratin | 10 |
| Ready meals & snacks | Potato products | Potato wedges & croquettes | 4 |
| Ready meals & snacks | Ready-to-eat food | Other dishes | 47 |
| Ready meals & snacks | Ready-to-eat food | Pancakes | 5 |
| Ready meals & snacks | Ready-to-eat food | Pies | 28 |
| Ready meals & snacks | Ready-to-eat food | Pizza | 9 |
| Ready meals & snacks | Ready-to-eat food | Porridge | 5 |
| Ready meals & snacks | Ready-to-eat food | Portion dishes | 27 |
| Ready meals & snacks | Ready-to-eat food | Pyttipanna | 2 |
| Ready meals & snacks | Ready-to-eat food | Soups | 22 |
| Ready meals & snacks | Snack | Bars | 9 |
| Ready meals & snacks | Snack | Dried fruit | 3 |
| Ready meals & snacks | Snack | Nuts | 6 |
| Ready meals & snacks | Snack | Sandwich | 15 |
| Ready meals & snacks | Snack | Sesame cookies | 4 |
| Ready meals & snacks | Vegetarian | Other vegetarian | 9 |
| Ready meals & snacks | Vegetarian | Tofu | 16 |
| Ready meals & snacks | Vegetarian | Vegetarian burgers | 2 |
| Ready meals & snacks | Vegetarian | Vegetarian sausage | 4 |
| Spices & Seasonings | Barbecuing | Barbecue sauces | 5 |
| Spices & Seasonings | Barbecuing | Glaze | 2 |
| Spices & Seasonings | Barbecuing | Marinade | 7 |
| Spices & Seasonings | Barbecuing | Rub | 2 |
| Spices & Seasonings | Broth & Stock | Broth | 8 |
| Spices & Seasonings | Ketchup, Mustard & Chili Sauce | Chili sauce | 5 |
| Spices & Seasonings | Ketchup, Mustard & Chili Sauce | Ketchup | 22 |
| Spices & Seasonings | Ketchup, Mustard & Chili Sauce | Mustard | 33 |
| Spices & Seasonings | Ketchup, Mustard & Chili Sauce | Tomato paste | 7 |
| Spices & Seasonings | Mayonnaise, Dressing & Other Flavorings | Dressing | 13 |
| Spices & Seasonings | Mayonnaise, Dressing & Other Flavorings | Mayonnaise | 18 |
| Spices & Seasonings | Mayonnaise, Dressing & Other Flavorings | Other flavourings | 2 |
| Spices & Seasonings | Oil & Vinegar | Vinegar | 5 |
| Spices & Seasonings | Sauces & Spice Butters | Dry sauce | 17 |
| Spices & Seasonings | Sauces & Spice Butters | Horseradish | 3 |
| Spices & Seasonings | Sauces & Spice Butters | Ready-made sauce | 47 |
| Spices & Seasonings | Sauces & Spice Butters | Spice butter | 4 |
| Spices & Seasonings | Spices & Herbs | Salt | 11 |
| Spices & Seasonings | Spices & Herbs | Spice Grinders | 6 |
| Spices & Seasonings | Spices & Herbs | Spices A - J | 56 |
| Spices & Seasonings | Spices & Herbs | Spices K - P | 50 |
| Spices & Seasonings | Spices & Herbs | Spices R - Z | 45 |

Appendix 2: Summary Statistics of Food Products from Swiss Supermarket

| Variable | Obs | Mean | Std. dev. | Min | Max |
| --- | --- | --- | --- | --- | --- |
| LnPrice | 5,329 | 1.40 | 0.56 | 0.00 | 3.74 |
| PackageSize | 5,329 | 335.91 | 363.03 | 0.35 | 5,000.00 |
| Climate | 2,947 | 3.50 | 1.25 | 1.00 | 5.00 |
| Carb | 4,915 | 96.73 | 161.67 | 0.00 | 4,000.00 |
| Fat | 4,291 | 33.97 | 44.86 | 0.00 | 710.00 |
| Protein | 4,917 | 27.86 | 34.31 | 0.00 | 480.00 |
| Salt | 4,906 | 3.93 | 25.09 | 0.00 | 998.00 |
| Energy | 4,909 | 817.63 | 846.79 | 0.00 | 16,000.00 |

Appendix 3: Food Products by Category from Swiss Supermarket

| Category Level 1 | Freq. | Percent | Cum. |
|---|---:|---:|---:|
| Bread, pastries & breakfast | 656 | 12.31 | 12.31 |
| Dairy, eggs & fresh convenience food | 939 | 17.62 | 29.93 |
| Frozen food | 397 | 7.45 | 37.38 |
| Fruits & vegetables | 296 | 5.55 | 42.93 |
| Meat & fish | 591 | 11.09 | 54.03 |
| Pasta, condiments & canned food | 1,233 | 23.14 | 77.16 |
| Snacks & sweets | 1,217 | 22.84 | 100 |
| Total | 5,329 | 100 | |

Appendix 4: Details on Categories of Food Products from Swiss Supermarket

| CategoryLevel1 | CategoryLevel2 | CategoryLevel3 | _freq |
|---|---|---|---|
| Bread, pastries & breakfast | Baking ingredients | Cake Dough & Pizza Dough | 42 |
| Bread, pastries & breakfast | Baking ingredients | Cooking chocolates | 11 |
| Bread, pastries & breakfast | Baking ingredients | Decorations & toppings | 36 |
| Bread, pastries & breakfast | Baking ingredients | Dessert mixes | 40 |
| Bread, pastries & breakfast | Baking ingredients | Flavours & gelling agents | 14 |
| Bread, pastries & breakfast | Baking ingredients | Flour & yeast | 40 |
| Bread, pastries & breakfast | Baking ingredients | Ground nuts | 15 |
| Bread, pastries & breakfast | Baking ingredients | Sugar & sweeteners | 25 |
| Bread, pastries & breakfast | Bread & rusk | Corn & rice pancakes | 22 |
| Bread, pastries & breakfast | Bread & rusk | Crispbread & rusk | 32 |
| Bread, pastries & breakfast | Bread & rusk | Packaged bread | 55 |
| Bread, pastries & breakfast | Bread & rusk | Pre-baked bread | 33 |
| Bread, pastries & breakfast | Bread & rusk | Toast | 15 |
| Bread, pastries & breakfast | Cereals & muesli | Cereals | 51 |
| Bread, pastries & breakfast | Cereals & muesli | Muesli | 59 |
| Bread, pastries & breakfast | Cereals & muesli | Oats & bran | 28 |
| Bread, pastries & breakfast | Fresh bread | Fresh bread | 25 |
| Bread, pastries & breakfast | Jam, spreads & honey | Honeys & molasses | 22 |
| Bread, pastries & breakfast | Jam, spreads & honey | Jams | 64 |
| Bread, pastries & breakfast | Jam, spreads & honey | Spreads | 23 |
| Bread, pastries & breakfast | Pastries | Fresh pastries | 4 |
| Dairy, eggs & fresh convenience food | Cheese | Fresh & spreadable cheeses | 34 |
| Dairy, eggs & fresh convenience food | Cheese | Goat & sheep cheese | 27 |
| Dairy, eggs & fresh convenience food | Cheese | Grated & sliced cheeses | 47 |
| Dairy, eggs & fresh convenience food | Cheese | Hard & semi-hard cheeses | 82 |
| Dairy, eggs & fresh convenience food | Cheese | Mozzarella, feta & cottage cheeses | 30 |
| Dairy, eggs & fresh convenience food | Cheese | Raclettes, fondues & grilling cheeses | 60 |
| Dairy, eggs & fresh convenience food | Cheese | Snacks, appetisers & cheese for kids | 29 |
| Dairy, eggs & fresh convenience food | Cheese | Soft cheeses & blue cheeses | 51 |
| Dairy, eggs & fresh convenience food | Fresh convenience food | Antipasti & sauces | 30 |
| Dairy, eggs & fresh convenience food | Fresh convenience food | Fresh pasta & lasagnes | 66 |
| Dairy, eggs & fresh convenience food | Fresh convenience food | Pizzas & tarte flambée | 21 |
| Dairy, eggs & fresh convenience food | Fresh convenience food | Ready-made meals | 46 |
| Dairy, eggs & fresh convenience food | Fresh convenience food | Ready-made salads | 21 |
| Dairy, eggs & fresh convenience food | Milk, butter & eggs | Butter & margarine | 34 |
| Dairy, eggs & fresh convenience food | Milk, butter & eggs | Creams & whipped cream | 12 |
| Dairy, eggs & fresh convenience food | Milk, butter & eggs | Eggs | 1 |
| Dairy, eggs & fresh convenience food | Milk, butter & eggs | Milk & dairy drinks | 4 |
| Dairy, eggs & fresh convenience food | Vegan alternatives to dairy products | Alternatives to yogurts & desserts | 24 |
| Dairy, eggs & fresh convenience food | Vegan alternatives to dairy products | Cheese alternatives | 12 |
| Dairy, eggs & fresh convenience food | Vegan alternatives to dairy products | Milk & cream alternatives | 1 |
| Dairy, eggs & fresh convenience food | Yogurts & desserts | Compots | 13 |
| Dairy, eggs & fresh convenience food | Yogurts & desserts | Creams & desserts | 66 |
| Dairy, eggs & fresh convenience food | Yogurts & desserts | Quarks & cream cheeses | 28 |
| Dairy, eggs & fresh convenience food | Yogurts & desserts | Yogurt drinks | 6 |
| Dairy, eggs & fresh convenience food | Yogurts & desserts | Yogurts | 175 |
| Dairy, eggs & fresh convenience food | Yogurts & desserts | Yogurts & kids' desserts | 19 |
| Frozen food | Convenience food | Delicacies & quiches | 23 |
| Frozen food | Convenience food | Fish & shellfish | 19 |
| Frozen food | Convenience food | Lasagne & pasta | 20 |
| Frozen food | Convenience food | Plant-based alternatives | 1 |
| Frozen food | Convenience food | Spring rolls | 8 |
| Frozen food | Fish & seafood | Breaded fish & fish fingers | 24 |
| Frozen food | Fish & seafood | Fish | 18 |
| Frozen food | Fish & seafood | Seafood & shellfish | 20 |
| Frozen food | Fruits & vegetables | Fruit | 29 |
| Frozen food | Fruits & vegetables | Vegetable mixes | 24 |
| Frozen food | Fruits & vegetables | Vegetables | 30 |
| Frozen food | Ice cream & desserts | Other ice creams | 2 |
| Frozen food | Ice cream & desserts | Pastries & desserts | 21 |
| Frozen food | Meat & poultry | Hamburger, meatballs & tartare | 10 |
| Frozen food | Meat & poultry | Meat for fondue chinoise | 2 |
| Frozen food | Meat & poultry | Other meats | 3 |
| Frozen food | Meat & poultry | Plant-based alternatives | 8 |
| Frozen food | Meat & poultry | Poultry & nuggets | 30 |
| Frozen food | Pizza, bread & apero pastries | Appetizers | 21 |
| Frozen food | Pizza, bread & apero pastries | Bread & pastries | 11 |
| Frozen food | Pizza, bread & apero pastries | Pizza & tarte flambée | 43 |
| Frozen food | Potatoes & chips | Chips & potatoes | 18 |
| Frozen food | Potatoes & chips | Other potato products | 12 |
| Fruits & vegetables | Fresh herbs & spices | Fresh herbs | 3 |
| Fruits & vegetables | Fresh herbs & spices | Garlic, onions & shallots | 10 |
| Fruits & vegetables | Fresh herbs & spices | Sprouts & shoots | 5 |
| Fruits & vegetables | Fruits | Apples & pears | 17 |
| Fruits & vegetables | Fruits | Bananas | 4 |
| Fruits & vegetables | Fruits | Citrus fruits | 14 |
| Fruits & vegetables | Fruits | Exotic fruits | 20 |
| Fruits & vegetables | Fruits | Grapes | 2 |
| Fruits & vegetables | Fruits | Red fruits & berries | 4 |
| Fruits & vegetables | Ready to use | Fruits | 8 |
| Fruits & vegetables | Ready to use | Vegetables | 23 |
| Fruits & vegetables | Root vegetables | Carrots & peas | 8 |
| Fruits & vegetables | Root vegetables | Celery & beetroot | 10 |
| Fruits & vegetables | Root vegetables | Potatoes | 22 |
| Fruits & vegetables | Root vegetables | Radish & fennel | 4 |
| Fruits & vegetables | Salad | Chicory | 6 |
| Fruits & vegetables | Salad | Packaged salads | 23 |
| Fruits & vegetables | Salad | Salads | 11 |
| Fruits & vegetables | Vegetables | Cauliflower, broccoli & cabbage | 22 |
| Fruits & vegetables | Vegetables | Courgettes, aubergines & peppers | 16 |
| Fruits & vegetables | Vegetables | Cucumbers & avocados | 2 |
| Fruits & vegetables | Vegetables | Green beans & sweetcorn | 10 |
| Fruits & vegetables | Vegetables | Mushrooms | 7 |
| Fruits & vegetables | Vegetables | Other vegetables | 14 |
| Fruits & vegetables | Vegetables | Tomatoes | 16 |
| Fruits & vegetables | Vitamin baskets | Fruits | 4 |
| Fruits & vegetables | Vitamin baskets | Fruits & vegetable mixes | 5 |
| Fruits & vegetables | Vitamin baskets | Vegetables | 6 |
| Meat & fish | Cold cuts | Bacon & diced bacon | 22 |
| Meat & fish | Cold cuts | Cooked, rolled & smoked ham | 51 |
| Meat & fish | Cold cuts | Deli meat & meatloaf | 14 |
| Meat & fish | Cold cuts | Deli poultry meat | 30 |
| Meat & fish | Cold cuts | Pâté, terrines & foie gras | 10 |
| Meat & fish | Cold cuts | Raw ham & cured meats | 56 |
| Meat & fish | Cold cuts | Salami & dried sausages | 66 |
| Meat & fish | Cold cuts | Sausages & cervelats | 41 |
| Meat & fish | Fish | Fresh fish & shellfish | 15 |
| Meat & fish | Fish | Salmon & smoked fish | 23 |
| Meat & fish | Fish | Shrimps | 9 |
| Meat & fish | Fish | Surimi & seafood delicacies | 14 |
| Meat & fish | Fish | Sushi | 6 |
| Meat & fish | Meat & poultry | Beef | 29 |
| Meat & fish | Meat & poultry | Chicken, turkey & duck | 46 |
| Meat & fish | Meat & poultry | Chinoise & charbonnade | 3 |
| Meat & fish | Meat & poultry | Minced meat & burger | 22 |
| Meat & fish | Meat & poultry | Ostrich & rabbit | 2 |
| Meat & fish | Meat & poultry | Pork | 37 |
| Meat & fish | Meat & poultry | Tartare, roast beef & carpaccio | 6 |
| Meat & fish | Meat & poultry | Veal, lamb & horse | 19 |
| Meat & fish | Plant-based alternatives | Alternatives to burgers, cutlets & falafel | 34 |
| Meat & fish | Plant-based alternatives | Alternatives to deli meat | 9 |
| Meat & fish | Plant-based alternatives | Alternatives to minced & diced meat | 11 |
| Meat & fish | Plant-based alternatives | Tofu | 16 |
| Pasta, condiments & canned food | Canned food & convenience food | Ready meals & ravioli | 76 |
| Pasta, condiments & canned food | Canned food & convenience food | Savoury spreads | 13 |
| Pasta, condiments & canned food | Canned food & convenience food | Tinned fish | 42 |
| Pasta, condiments & canned food | Canned food & convenience food | Tinned fruit | 44 |
| Pasta, condiments & canned food | Canned food & convenience food | Tinned meat | 17 |
| Pasta, condiments & canned food | Canned food & convenience food | Tinned mushrooms | 18 |
| Pasta, condiments & canned food | Canned food & convenience food | Tinned tomatoes | 31 |
| Pasta, condiments & canned food | Canned food & convenience food | Tinned vegetables | 58 |
| Pasta, condiments & canned food | International food | Africa & the Orient | 6 |
| Pasta, condiments & canned food | International food | Asia | 71 |
| Pasta, condiments & canned food | International food | Korea | 11 |
| Pasta, condiments & canned food | International food | Mexico | 48 |
| Pasta, condiments & canned food | Pasta, rice, semolina & grain | Mashed potatoes & rösti | 18 |
| Pasta, condiments & canned food | Pasta, rice, semolina & grain | Pasta | 172 |
| Pasta, condiments & canned food | Pasta, rice, semolina & grain | Pulses & cereals | 67 |
| Pasta, condiments & canned food | Pasta, rice, semolina & grain | Rice | 47 |
| Pasta, condiments & canned food | Soups & stock | Bases & sauces for binding | 7 |
| Pasta, condiments & canned food | Soups & stock | Soups, potages & croutons | 52 |
| Pasta, condiments & canned food | Soups & stock | Stock cubes | 46 |
| Pasta, condiments & canned food | Spices & sauces | Cold sauces | 58 |
| Pasta, condiments & canned food | Spices & sauces | Cooking sauces | 35 |
| Pasta, condiments & canned food | Spices & sauces | Oil & vinegar | 2 |
| Pasta, condiments & canned food | Spices & sauces | Other ingredients | 11 |
| Pasta, condiments & canned food | Spices & sauces | Pasta sauces & pesto | 74 |
| Pasta, condiments & canned food | Spices & sauces | Pickled vegetables | 51 |
| Pasta, condiments & canned food | Spices & sauces | Salt & pepper | 27 |
| Pasta, condiments & canned food | Spices & sauces | Spices, herbs & seasonings | 131 |
| Snacks & sweets | Biscuits | Assorted biscuits | 4 |
| Snacks & sweets | Biscuits | Biscuits & sponge fingers | 46 |
| Snacks & sweets | Biscuits | Chocolate biscuits & cookies | 60 |
| Snacks & sweets | Biscuits | Filled cookies | 22 |
| Snacks & sweets | Biscuits | Fruit biscuits | 13 |
| Snacks & sweets | Biscuits | Meringues & bricelet wafers | 4 |
| Snacks & sweets | Biscuits | Shortbread & macaroons | 22 |
| Snacks & sweets | Biscuits | Wafers & läckerli | 13 |
| Snacks & sweets | Cake, madeleines & panettone | Cakes & dry cakes | 23 |
| Snacks & sweets | Cake, madeleines & panettone | Madeleines & small cakes | 28 |
| Snacks & sweets | Cake, madeleines & panettone | Panettone | 6 |
| Snacks & sweets | Cake, madeleines & panettone | Waffles and pancakes | 6 |
| Snacks & sweets | Chocolate & sweets | Bars, branches & snacks | 153 |
| Snacks & sweets | Chocolate & sweets | Bonbons | 108 |
| Snacks & sweets | Chocolate & sweets | Caramels & nougats | 6 |
| Snacks & sweets | Chocolate & sweets | Chewing gum | 27 |
| Snacks & sweets | Chocolate & sweets | Chocolate bars | 168 |
| Snacks & sweets | Chocolate & sweets | Pralines | 50 |
| Snacks & sweets | Crackers, cereal & fruit bars | Cereal bars | 41 |
| Snacks & sweets | Crackers, cereal & fruit bars | Crackers | 55 |
| Snacks & sweets | Crackers, cereal & fruit bars | Fruit bars | 14 |
| Snacks & sweets | Dried fruit | Dried fruit | 41 |
| Snacks & sweets | Dried fruit | Mixes | 18 |
| Snacks & sweets | Dried fruit | Nuts & almonds | 24 |
| Snacks & sweets | Snacks & crisps | Crisps | 94 |
| Snacks & sweets | Snacks & crisps | Dips & sauces | 11 |
| Snacks & sweets | Snacks & crisps | Olives & antipasti | 42 |
| Snacks & sweets | Snacks & crisps | Peanuts & salted nuts | 45 |
| Snacks & sweets | Snacks & crisps | Popcorn | 7 |
| Snacks & sweets | Snacks & crisps | Savoury biscuits | 66 |